\documentclass[aps,prd,reprint,superscriptaddress, longbibliography]{revtex4-1}

\usepackage{amsmath,graphicx,units,float,dcolumn}
\usepackage[colorlinks=true,allcolors=blue]{hyperref}

\begin{document}
\title{Joint cosmological inference of standard sirens and gravitational wave weak lensing}

\author{Giuseppe Congedo}
\email{giuseppe.congedo@ed.ac.uk}
%\affiliation{Institute for Astronomy, School of Physics and Astronomy, 
%University of Edinburgh, Royal Observatory, Blackford Hill, 
%Edinburgh, EH9 3HJ, United Kingdom}

\author{Andy Taylor}

\affiliation{Institute for Astronomy, School of Physics and Astronomy, 
University of Edinburgh, Royal Observatory, Blackford Hill, 
Edinburgh, EH9 3HJ, United Kingdom}

\date{\today}

\begin{abstract}

We present the first joint inference of standard sirens and 
gravitational wave weak lensing by filtering of the same dataset.
We imagine a post-LISA scenario emerging around the late 2030s when LISA will have accumulated a number of detections at high redshift;
LIGO-VIRGO will have finished observing at low redshift, and Einstein Telescope will have started making new observations out to redshifts possibly overlapping with LISA.
Euclid and other cosmological probes will have provided constraints at percent level by then, but mostly exhausted their ability to improve any further.
We derive forecasts assuming $\sim1\,\text{deg}^{-2}$ detected sources,
in conjunction with a spectroscopic follow-up (e.g.\ by Euclid, DESI, or ATHENA).
Thanks to the statistical power of standard sirens as a geometry probe
-- lifting key degeneracies in the gravitational wave weak lensing --
and no external priors assumed,
the constraints on dark matter and its clustering, namely $\Omega_m$ and $\sigma_8$, could be achieved to $2\%$ and $3\%$.
The Hubble constant could be constrained to better than $1\%$ in all cases;
the dark energy density, $\Omega_\Lambda$, to $2\%$, and curvature, $\Omega_K$, to $0.02$;
the amplitude and spectral tilt of the scalar fluctuations, $\ln(10^{10}A_s)$ and $n_s$, to $2\%$ and $7\%$.
As a completely independent cosmological probe,
with less calibration requirements,
the joint inference of standard sirens and gravitational wave weak lensing might help solve the tensions
currently observed between other cosmological probes,
such as CMB, galaxy lensing, and Type Ia SNs,
and distinguish between residual systematics and new physics.

\end{abstract}

%\pacs{}
%\keywords{}

\maketitle

%%%%%%%%%%%%%%%%%%%%%%%%%%%%%%%%

\section{Introduction}

The gravitational wave (GW) distance-redshift relation can be exploited to derive independent cosmological constraints on $H_0$,
as demonstrated by the LIGO-VIRGO (LV) collaboration with 
the first multi-messenger event \cite{abbott2017}.
Moreover, GW measurements can now be combined with Planck \cite{planck2018a} to further improve constraints on other parameters,
such as neutrino mass, curvature, and dark energy equation of state \cite{divalentino2018}.
In future, we will have collected a plethora of GW events,
attained from the combination of different source catalogues,
namely by LV, Einstein Telescope (ET) \cite{punturo2010}, and LISA \cite{amaro2017}, reaching median redshifts of $\sim$2.
Precious cosmological information will then be extracted by having access to luminosity distance, redshift, and sky position data.

In constrast to Type Ia supernovae (SN) analysis which is subjected to external calibration (the cosmic distance ladder) \cite{riess2016,riess2018},
the distance to the GW source can instead be measured directly and accurately with GW data alone \cite{schutz1986}.
Owing to the degeneracy between mass and redshift, the corresponding redshift must however be derived from electromagnetic counterparts through multi-messenger events.
The weak lensing (WL) from large-scale structure (LSS) is also known
to affect the distance-redshift relation  \cite{holz2005}.
The farther the sources are, the bigger the effect will be,
hence the bigger the scatter of the luminosity distance being randomly magnified/demagnified by over/under matter densities.
This percent level effect is comparable with the GW measurement error and
has traditionally been seen as a source of potential systematic that could in principle be corrected for \cite{hilbert2011,hirata2010},
or more recently as an actual cosmological probe on its own for future, more dedicated, GW experiments \cite{camera2013,cutler2009}.

Just as the cosmic microwave background (CMB) and the lensing of the CMB jointly provide additional constraining power
for cosmogical inference, as demonstrated by the Planck satellite \cite{planck2018b},
in this paper we propose a joint analysis of standard sirens and weak lensing by filtering of the same GW data,
such that
the weak lensing inference effectively becomes conditional to the standard sirens data.
We derive the first joint cosmological forecasts for data as it will be
observed post the space-based detector LISA in the late 2030s,
in conjunction with other detectors (e.g.\ LV and ET),
and redshifts from spectroscopic follow-up surveys such as Euclid \cite{laureijs2011}, DESI \cite{desi2016}, or ATHENA \cite{athena2014}.
We show how the method will provide competitive constraints, 
which will be alternative and complementary to galaxy surveys and CMB experiments.
By that time the same experiments will have exhausted their 
ability to improve their cosmological constraints significantly below the percent level as it becomes more and more apparent in the era of systematics-dominated cosmology.

Throughout the rest of the paper, we assume the 2018 Planck cosmology, $H_0=\unit[100\,h]{km\,s^{-1} Mpc^{-1}}$, $h=0.673$, $h^2\Omega_m=0.143$, $h^2\Omega_\Lambda=0.310$, $\Omega_K=0$, $\ln(10^{10} A_s)=3.05$, and $n_s=0.965$ \cite{planck2018a}.

%%%%%%%%%%%%%%%%%%%%%%%%%%%%%%%%

\section{Standard sirens and weak lensing}

Interferometric detectors are sensitive to the derivative of the light frequency shift induced by incoming GWs \cite{congedo2015,congedo2013}.
Owing to amplitude and frequency modulations induced by the motion of the detector around the Sun (LISA) or Earth (LV and ET), the amplitude of the wave is particularly well measured.
As the observed GW strain is inversely proportional to the luminosity distance \cite{congedo2017,maggiore2009},
the luminosity distance is determined very accurately.

The standard sirens inference is based on relating the GW luminosity distance, $d_L$,
to the optical redshift, $z$, as follows
\begin{equation}
d_L(z)=(1+z)f_K(\chi),
\end{equation}
where the comoving angular distance is given by
\begin{equation}
f_K(\chi) = \frac{1}{\sqrt{-K}}\sinh\left[\sqrt{-K}\chi\right].
\end{equation}
In the above equation, $K$ is the curvature and the comoving distance is
\begin{equation}
\chi(z)=c\int_0^z\frac{\text{d}z'}{H(z')},
\end{equation}
where $H(z)=H_0[\Omega_m(1+z)^3+\Omega_K(1+z)^2+\Omega_\Lambda]^{1/2}$ and $\Omega_K=-(c/H_0)^2 K$.

With a nominal sky resolution of $\sim\unit[1]{deg^2}$, it would, in principle, be possible to
determine the redshift for the detected source,
either from an electromagnetic counterpart, or an ensemble average over a population of galaxies selected in the same sky bin.
Potentially, the redshift could also be inferred directly from GW data
(see e.g.\ Ref.\;\cite{bonvin2017} and references therein).
Given that the redshift selection function of GW observations will be required to have good overlap with optical follow-up surveys,
it is reasonable to assume in our analysis that
(\textit{i}) the electromagnetic counterpart can indeed be identified in the provided position error box;
(\textit{ii}) the redshift can be determined by a spectroscopic follow-up survey, which is feasible given the relative small number of GW sources involved in this analysis compared to galaxy surveys.

We adopt the following scaling relation for the expected total redshift error,
\begin{equation}\label{eq:redshift_covariance}
\sigma^2_z=\left(\sigma_{\text{sp}}^2+\frac{\sigma^2_v}{c^2}\right)(1+z)^2,
\end{equation}
which accounts for both the spectroscopic error $\sigma_{\text{sp}}=0.001$ as per the Euclid/DESI nominal requirements,
and the peculiar velocity dispersion, being $\sigma_v\sim\unit[300]{km/s}$ a reasonable estimate
\footnote{We note however that the peculiar motion can affect the total redshift error budget only for very shallow observations,
whereas it becomes totally negligible at, e.g., the LISA depth.}.

Current forecasts for LISA \cite{tamanini2017,tamanini2016} are based only on the $d_L$ versus $z$ relation with reasonable expectations of measurement errors similarly to the above.
However, the forecasts for models beyond flat $\Lambda$CDM are usually derived by setting key parameters (e.g.\ $h$) to nominal values,
instead of correctly marginalising over their probability distribution,
and the constraints on the dark sector are generally not very strong.

As noted in Ref.\;\cite{cutler2009}, the WL fluctuations can be directly inferred from GW data.
The WL magnification factor is defined as
\begin{equation}
\mu=\frac{1}{(1-\kappa)^2-|\gamma|^2}\sim1+2\kappa,
\end{equation}
where $\kappa$ and $\gamma$ are the convergence and shear fields owing to the lensing by the LSS, and the approximation holds in the weak regime, $\kappa,|\gamma|\ll1$.
The magnification factor describes how solid angles, 
hence angular diameter distances, are transformed 
under lensing and is proportional, to leading order, to the convergence field.
Likewise, the luminosity distance is transformed as follows
\begin{equation}\label{eq:weak_lensing_distance}
d_L'=(1-\kappa)d_L,
\end{equation}
where primed denotes a lensed quantity.
Note that $d_L'$ is the only observed quantity, as the detected GW strain is $h\propto1/d_L'$.
In effect, the GW luminosity is $L\propto \dot{h}^2=f^2/{d_L'}^2$, where $f$ is the frequency of the wave.
However, in general, $d_L$ is also affected by, e.g.\ cosmological and gravitational redshift \cite{yoo2016}.
Similarily, the phase of the GW signal suffers from analogous deviations \cite{bonvin2017}.
These effects are not included in our analysis.
They may well be treated as systematic errors or even, potentially, as source of extra information \cite{bonvin2006}.

As the rms of the $\kappa$ field is $\sim0.05$ at $z\sim2$,
this induces a typical lensing error on $d_L$ of $5\%$
(incidentally, this is comparable with or bigger than the GW
measurement error on $d_L$ that LISA will be able to achieve, as discussed later on in this section).
We adopt, however, the more realistic fitting formula of Refs \cite{tamanini2016,hirata2010}, which predicts the lensing fluctuation on $d_L$, as a function of redshift.

We can now derive an estimate of the GW-WL convergence by adopting a fiducial model for the unlensed $d_L$,
and inverting Eq.\;\eqref{eq:weak_lensing_distance}.
In fact, assuming a typical measurement error on $d_L$,
we would easily get a point estimate of $\kappa$ of 
significance $S/N>1$ with a single GW detection,
as opposed to optical WL
where an ensemble average of $\sim10^3$ identical sources or more are required to get the same statistical power.
The consequent reduction in the effective survey sample is therefore $\sim10^3$.

Although noisy and conditional to the measured $d_L$,
the GW estimate of $\kappa$ would be sufficient to calculate
a 2D power spectrum. We define the convergence power spectrum as follows \cite{kilbinger2015}
\begin{equation}
\mathcal{C}^\kappa_\ell=\int_0^{\chi_d}\text{d}\chi W^2(\chi)\mathcal{P}_\delta\left(k=\frac{\ell+1/2}{f_K(\chi)}, \chi\right),
\end{equation}
where $\mathcal{P}_\delta(k,\chi)$ is the 3D power spectrum of matter density fluctuations,
$k$ is the Fourier mode, $\ell$ is the spherical harmonic multipole, $\chi_d=\chi(z_d)$, and $z_d$ is the survey depth.
The weight function defining the lensing efficiency is given by
\begin{equation}
W(\chi)=\frac{3}{2}\left(\frac{H_0}{c}\right)^2\frac{\Omega_m}{a(\chi)}\int_{\chi}^{\chi_d}\text{d}\chi'n(\chi')\frac{f_K(\chi'-\chi)}{f_K(\chi')},
\end{equation}
where $a(\chi)$ is the scale factor, $n(\chi)=n(z)|d\chi/dz|^{-1}$ is the source redshift distribution, and $\bar{n}=\int n(z)\text{d}z$ is the survey mean number density per steradian.

The lowest multipole accessible with a WL analysis of GW observations is limited by cosmic variance,
which is determined by the sky coverage of the optical survey, say $\ell_{\min}\sim2$.
The highest multipole is instead
limited by shot noise, which is given by the GW angular resolution. Assuming a realistic error box of $\sim\unit[1]{deg^{-2}}$,
that brings us to $\ell_{\max}\sim100$,
which is in contrast to other WL forecasts that assume
$\ell_{\max}\sim1,000$ or more.
Therefore a WL analysis of GW data would only probe the large scales,
and not be sensitive to the fully non-linear scales typical of galaxy WL surveys.
In general, the total GW-WL noise variance is given by
\begin{equation}\label{eq:weak_lensing_covariance}
\Sigma^\kappa_{\ell\ell}=\frac{2}{f_{\text{sky}}(2\ell+1)\delta\ell}
\left(\mathcal{C}^\kappa_\ell+\frac{\sigma^2_\kappa}{\bar{n}}\right)^2,
\end{equation}
where $\delta\ell$ is the multipole resolution,
$f_{\text{sky}}\sim0.4$ is the sky coverage that would be attained by an optical follow-up by a galaxy survey like Euclid,
$\sigma_\kappa$ is the total rms error on our estimate of convergence.
The first term is the cosmic variance and the second term is the shot noise.
Because the shot noise variance is $\propto\sigma_\kappa^4/\bar{n}^2$ and $\bar{n}$ is expected be small, GW-WL will in general be shot noise dominated.
In this case, the total signal-to-noise ($S/N$) will be $\propto\bar{n}$.

We assume a nominal $\sim1\%$ error on luminosity distance
from GW detection.
To put this figure into context, the LV multi-messenger event had 
an error of $\sim15\%$ \cite{abbott2017}, whereas this is 
expected be around $1\%$ for stellar binary black holes 
\cite{kyutoku2017} and $5\%$ for extreme mass ratio inspirals
\cite{babak2017}, both in the LISA band.
Supermassive binary black holes will likely be observed in 
a greater number and with a much higher $S/N$,
hence resulting in a similar or even better precision.
We include spectroscopic redshift errors and peculiar velocities as they also
contribute to the total error budget, but are in general subdominant.
We adopt an $n(z)$ selection function as customary 
in galaxy surveys (see e.g.\ Ref.\;\cite{laureijs2011}), but with a peak at $z\sim2$, which agrees with predictions for e.g.\ LISA \cite{tamanini2017}.
We therefore estimate that a conservative number density $\bar{n}\sim\unit[1]{deg^{-2}}$
would be enough to allow the detection of the GW-WL signal at a $S/N\sim20$ significance level.

\section{Joint Fisher forecast}

We adopt a Bayesian approach to combine
the standard sirens and WL inference.
We wish to derive the cosmological posterior probability given
joint measurements of $d_L$, $\kappa$, and $\mathcal{C}^\kappa_\ell$, call this $p(\theta|\mathcal{C}^\kappa_\ell,\kappa,d_L)$,
where $\theta$ is the set of cosmological parameters.
Thanks to the Bayes theorem, this probability is $\propto p(\mathcal{C}^\kappa_\ell,\kappa,d_L|\theta)\,p(\theta)$,
where the former is the joint likelihood and the latter is the cosmological prior.
By applying the conditional probability twice, the joint likelihood becomes
\begin{equation}
\begin{split}
p(\mathcal{C}^\kappa_\ell,\kappa,d_L|\theta) &=
p(\mathcal{C}^\kappa_\ell,\kappa|d_L,\theta)\,p(d_L|\theta) \\
&=p(\mathcal{C}^\kappa_\ell|\kappa,d_L,\theta)\,p(\kappa|d_L,\theta)\,p(d_L|\theta).
\end{split}
\end{equation}
In the first line, $p(\mathcal{C}^\kappa_\ell,\kappa|d_L,\theta)$ is the joint likelihood of power spectrum and convergence field conditional to standard sirens;
$p(d_L|\theta)$ is the standard sirens likelihood.
Upon further applying the conditional probability in the second line,
$p(\mathcal{C}^\kappa_\ell|\kappa,d_L,\theta)$ becomes the power spectrum likelihood conditional on both standard sirens and convergence field;
finally, $p(\kappa|d_L,\theta)$ is the convergence field likelihood conditional on standard sirens.
The equation shows how a joint likelihood for power spectrum, convergence field and standard sirens can be derived.
For simplicity, we will not, however, model $p(\kappa|d_L,\theta)$ in our analysis, and defer that to future work.
We just note that, as we are effectively restricted to the linear regime ($\ell_{\max}\sim100$), 
a measure of the variance of the convergence map as constructed above can be used to constrain the amplitude of the clustering, 
through $\sigma_8$ (see e.g.\ Ref.\;\cite{patton2017}).
Moreover, $\mathcal{C}^\kappa_\ell$ and $\kappa$ (via $d_L$) are in 
general correlated. It turns out that this happens at the level of 
a 3-point statistic (or bi-spectrum),
which may be significant only at highly non-linear scales.
Once again, because we work in the linear regime, 
it is safe to ignore this correlation and assume that individual 
Fisher matrices can be safely summed up together.

For the purpose of this investigation,
we define two baseline cosmologies: (\textit{a}) the concordance flat $\Lambda$CDM model with free parameters $\theta=\{h,h^2\Omega_m,\ln(10^{10} A_s),n_s\}$ and constraint $\Omega_K=0$; 
(\textit{b}) the extended $\Lambda$CDM model with curvature, i.e.\ $\theta=\{h,h^2\Omega_m,h^2\Omega_K,\ln(10^{10} A_s),n_s\}$.
In either cases, $h^2\Omega_\Lambda$ and $\sigma_8$,
the amplitude of the linear matter power spectrum at $\unit[8]{h^{-1}Mpc}$, are derived parameters.
In particular, $\sigma_8$ correlates strongly with $A_s$ and $n_s$, 
the amplitude and spectral tilt of the primordial scalar fluctuations.
We note that standard sirens are a geometry probe, 
in that their parameter space is restricted to 
$\theta=\{h,h^2\Omega_m,h^2\Omega_K\}$ and therefore this does not bring any information about the clustering.

We wrote \texttt{Python} \cite{python} code to derive our forecasts.
We calculate the luminosity distance and its derivatives with respect to cosmology semi-analytically,
whereas we get the power spectrum from \texttt{CLASS} \cite{blas2011},
whose numerical derivatives are robustly estimated with the \texttt{numdifftools}  package \cite{numdifftools}.
We assume that the likelihoods are Gaussian in their data, and all the measurement errors are Gaussian and uncorrelated.
In particular, we assume the WL noise covariance in Eq.\,\eqref{eq:weak_lensing_covariance}, 
and nominal GW errors of $1\%$ on $d_L$, plus redshift errors, peculiar velocities [see Eq.\,\eqref{eq:redshift_covariance} and text thereafter], and lensing errors for standard sirens.
We then calculate the individual Fisher matrices for standard sirens and WL,
and the Fisher matrix for the corresponding joint analysis
as the sum of the two.
These Fisher matrices are derived for the physical parameters, $h^2\Omega_i$, as customary for standard cosmological measurements,
and then mapped to density parameters, $\Omega_i$.
To do that, we first draw samples from the initial Fisher matrices in the physical parameter space,
and then map those samples to the density space via their non-linear transformation.
In doing so, the following flat priors are assumed: $0\le h\le1$, $0\le \Omega_m\le 1+\Omega_K$, $|\Omega_K|\le 0.3$, $2\le \ln(10^{10}A_s)\le 4$, and $0.5\le n_s\le 1.5$.
This general procedure has the advantage that it can reproduce the typical degeneracies that are seen between, e.g., $\Omega_m$ and $\sigma_8$ when the noise is large.
At the same time this procedure does not affect the final Fisher matrix estimation in any case, and can also be applied to any n-dimensional cosmologies.
It does however require a Monte Carlo simulation over multiple cosmologies, hence calling the power spectrum calculation multiple times to compute derived parameters such as $\sigma_8$.

\section{Results and discussion}

We constructed 2D marginal contour plots of the relevant parameters,
whose samples were drawn with a Monte Carlo simulation,
following the procedure described in the previous section.
In what follows, shown are one and two sigma contour plots for individual and 
combined analyses for the curved cosmology.
For the sake of completeness, results are also summarised in Table\;\ref{tab:results}
where 1D marginal errors are reported for both cosmologies and individual probes ($d_L$, $\mathcal{C}^\kappa_\ell$, and jointly).

\begin{table*}[!htbp]
\caption{Forecasted 1D marginal constraints for flat $\Lambda$CDM (left columns) and $\Lambda$CDM with curvature (right columns).
Shown are fractional percent errors for each of the relevant cosmological parameters that are considered in this paper.
Standard sirens, $d_L$, bring in the best constraint on geometry ($h$ and $\Omega_i$),
but do not say anything about clustering. However, when combined with gravitational wave weak lensing, $\mathcal{C}^\kappa_\ell$,
clustering ($\sigma_8$, $A_s$, and $n_s$) can also be improved to percent level.}
\label{tab:results}
\begin{ruledtabular}
\begin{tabular}{r d d d d d d d d d d d d}
& \multicolumn{2}{c}{$h$} & \multicolumn{2}{c}{$\Omega_m$} & \multicolumn{2}{c}{$\sigma_8$} & \multicolumn{2}{c}{$\ln(10^{10}A_s)$} & \multicolumn{2}{c}{$n_s$} & \multicolumn{1}{c}{$\Omega_\Lambda$} & \multicolumn{1}{c}{$\Omega_K$} \\
\hline
$d_L$ & 0.21 & 1.1 & 1.2 & 2.4 & \text{-} & \text{-} & \text{-} & \text{-} & \text{-} & \text{-} & 2.0 & 0.020 \\
$\mathcal{C}^\kappa_\ell$ & 77 & 390 & 61 & 44 & 27 & 23 & 22 & 290 & 34 & 130 & 26 & 0.16 \\
$d_L$ + $\mathcal{C}^\kappa_\ell$ & 0.21 & 1.1 & 1.1 & 2.4 & 2.7 & 2.7 & 1.9 & 2.2 & 6.7 & 6.7 & 2.0 & 0.019
\end{tabular}
\end{ruledtabular}
\end{table*}

\begin{figure*}[!htbp]
\centering
\begin{tabular}{cc}
\vspace{-5pt} \includegraphics[width=\columnwidth]{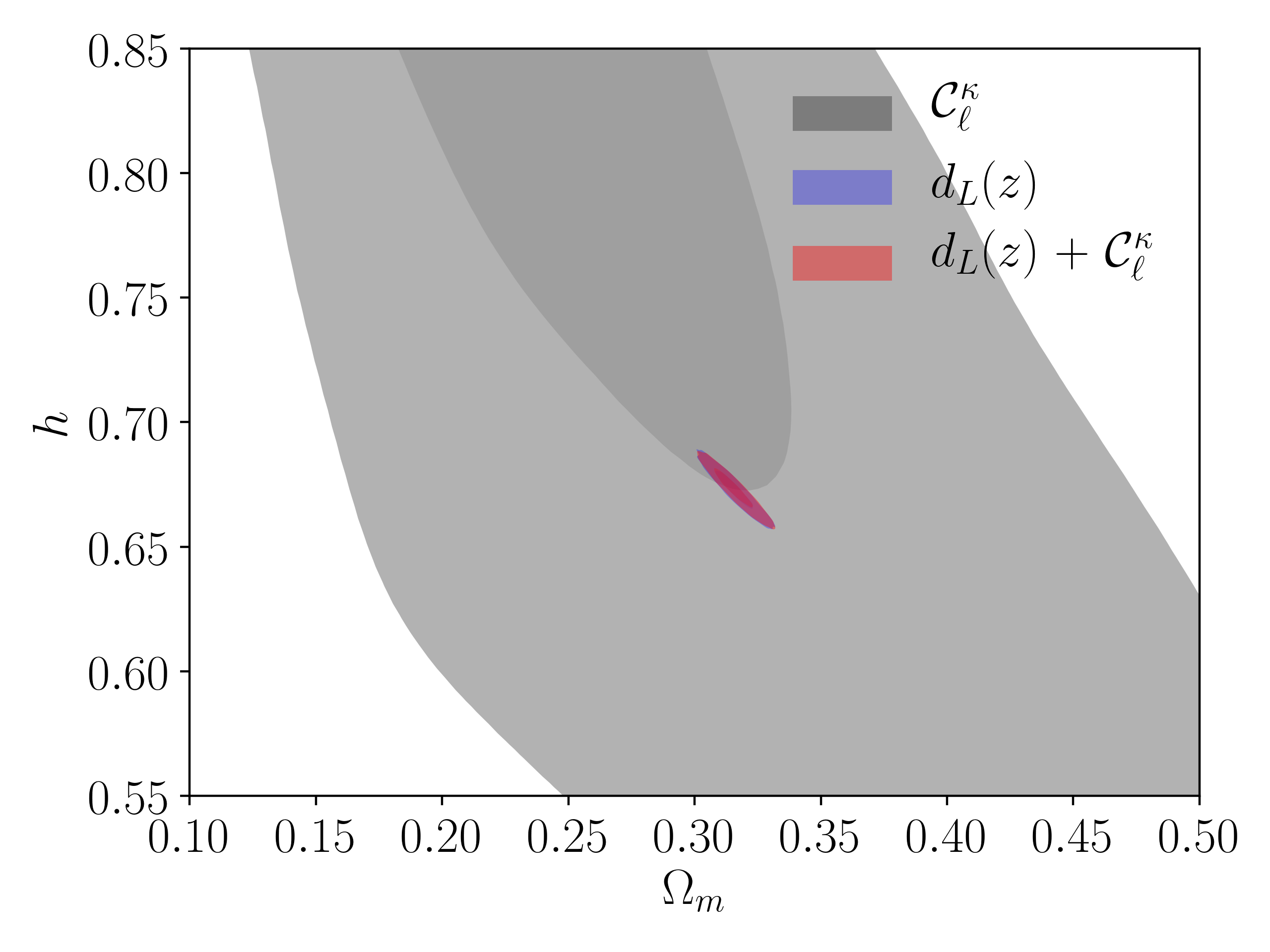} & \includegraphics[width=\columnwidth]{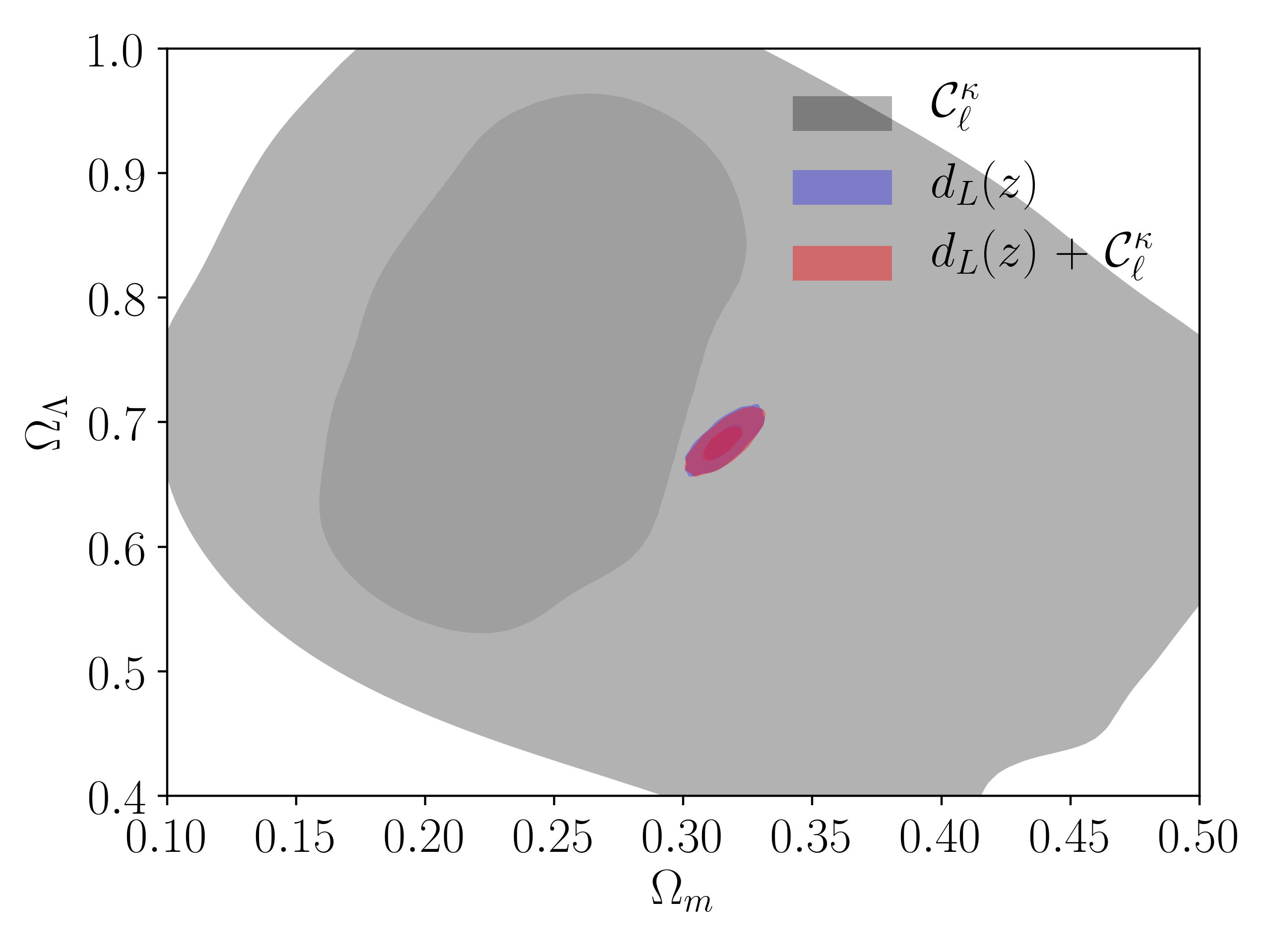} \\
\qquad \footnotesize{(a)} & \qquad \footnotesize{(b)} \\
\vspace{-5pt} \includegraphics[width=\columnwidth]{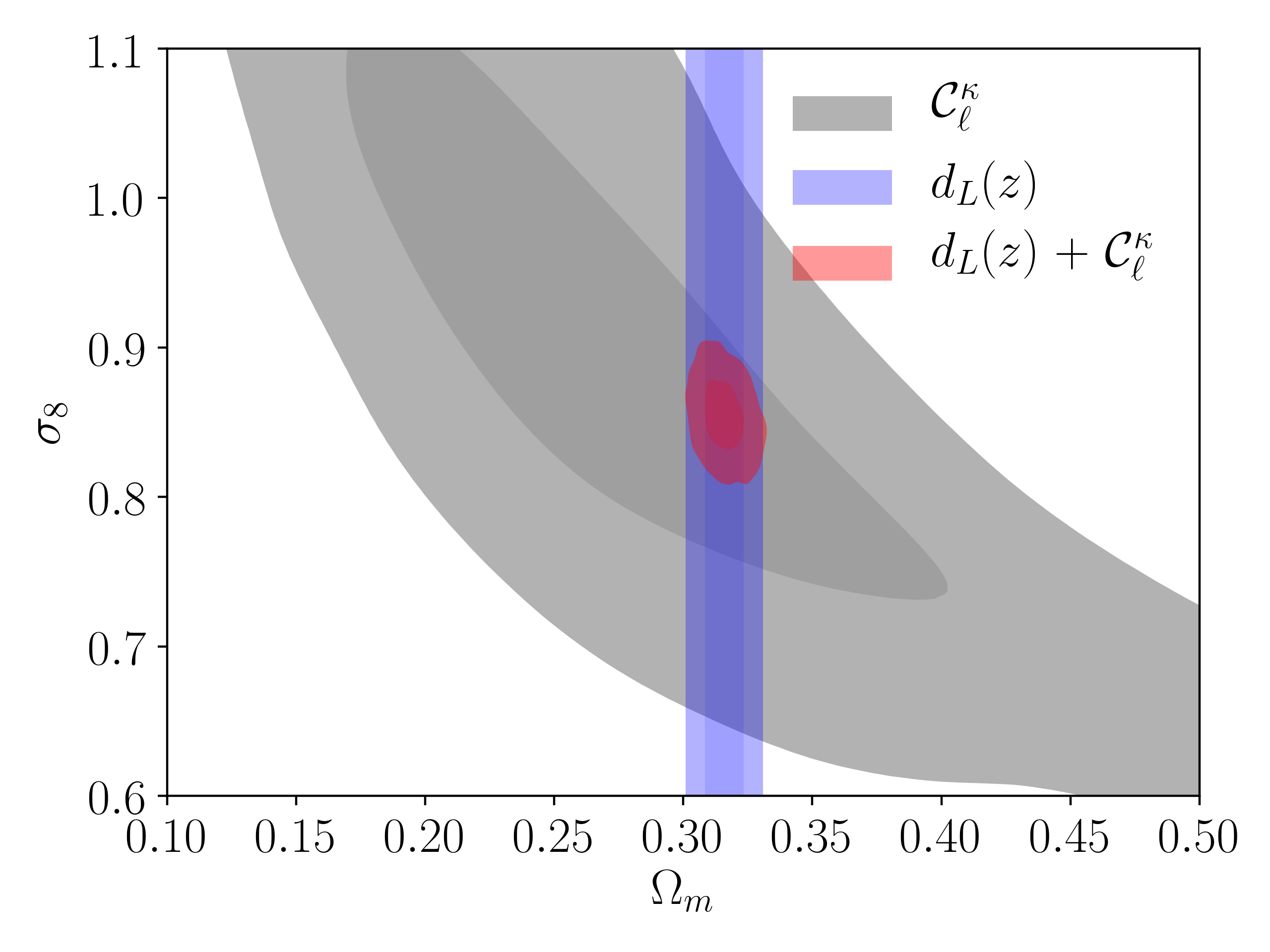} & \includegraphics[width=\columnwidth]{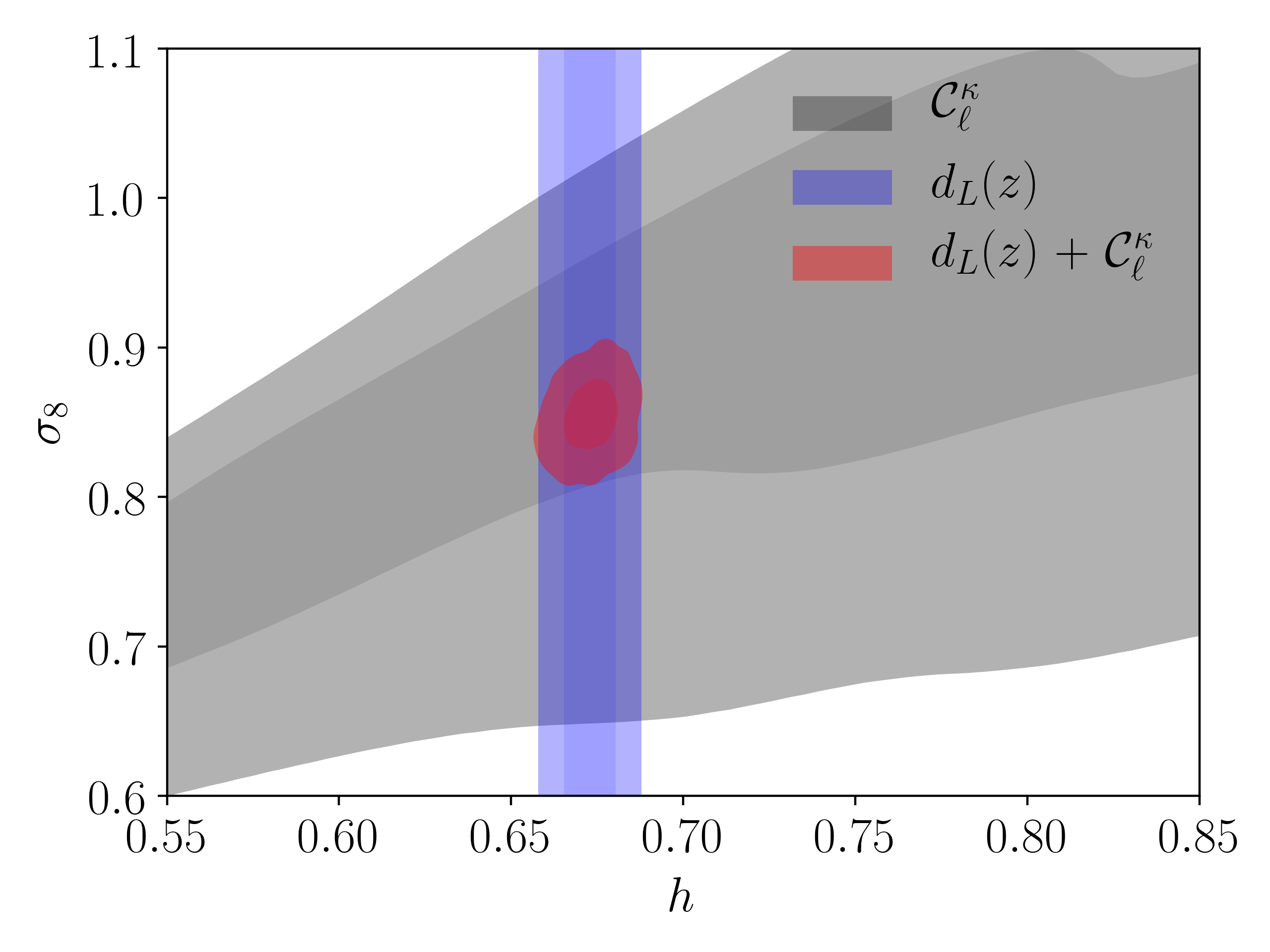} \\
\qquad \footnotesize{(c)} & \qquad \footnotesize{(d)}
\end{tabular}
\caption{\footnotesize{Forecasted 2D marginal constraints for $\Lambda$CDM with curvature:
gravitational wave weak lensing, $\mathcal{C}_\ell^\kappa$;
standard sirens, $d_L$; and jointly, $d_L$ + $\mathcal{C}_\ell^\kappa$.
As a geometry probe, standard sirens bring in most of the information [see panels (a) and (b)].
Although they cannot constrain clustering on their own,
they do however break the weak lensing degeneracy [see negative and positive slope of $\mathcal{C}_\ell^\kappa$ in panels (c) and (d)],
and help bring the error for clustering down to percent level.
Therefore clustering is very well determined by the combination of the two probes.
The projected joint errors (see also Table\;\ref{tab:results} for reference) are:
$1.1\%$ on $h$, $2.4\%$ on $\Omega_m$, $2\%$ on $\Omega_\Lambda$, and $2.7\%$ on $\sigma_8$.
Compared to flat $\Lambda$CDM, here the dilation of the errors owing to the inclusion of curvature is e.g.\ a factor 2 for $h$ and 5 for $\Omega_m$.}}
\label{fig:contours}
\end{figure*}

We start from the $\Omega_m$-$h$ contour plot [see Fig.\;\ref{fig:contours}, panel (a)].
We note that because of shot noise at such a small number density,
the WL constraints are generally poorer than standard sirens.
Also, because of noise
and the non-linear mapping from physical parameters to density parameters, the WL contour emerges as slightly shifted (i.e.\ biased) at $\sim0.5\,\sigma$ level from the nominal cosmology.
This is a noise bias effect that is absolutely expected here
given the low $S/N$ of the WL observable.
We checked that this is automatically reduced with bigger number densities, and therefore higher $S/N$.
By quantifying this error, one could also think about correcting it
in the first place.
In the shot noise limit, the information on both parameters is primarily driven by standard sirens.
The joint error on $h$ is $0.21\%$ for flat cosmology, degraded to $1.1\%$ with curvature.
Similarly, the error on $\Omega_m$ is $1.1\%$, degraded to $2.4\%$.

The second contour plot, $\Omega_m$-$\Omega_\Lambda$ [see Fig.\;\ref{fig:contours}, panel (b)]
is the primary constraint on dark energy.
Here we find a situation that is very similar to $\Omega_m$-$h$ -- as a geometry probe standard sirens drive most of the information,
and the constraint is much better than that of WL.
We predict a joint error on $\Omega_\Lambda$ of $2\%$ from $26\%$ with only WL.
Our constraint on $\Omega_K$ is $\sim0.02$.

The third and fourth contour plots, $\Omega_m$-$\sigma_8$ and $h$-$\sigma_8$ [see Fig.\;\ref{fig:contours}, panels (c) and (d)] -- the main result of this paper -- illustrate the benefit of
combining standard sirens with GW-WL,
which is sensitive to clustering.
Here the constraint from WL alone would already be competitive with galaxy WL surveys to date,
even without applying any informative prior on $h$ (apart hard bounds such as $0\le h\le1$)
as instead done in galaxy surveys (see e.g.\ Refs \cite{abbott2018,kohlinger2017}).
By combining with standard sirens -- a geometry probe with great statistical power on $h$ and $\Omega_m$ --
the $\Omega_m$-$\sigma_8$ as well as $h$-$\sigma_8$ degeneracies are broken and the constraint on clustering is improved dramatically.
Note that the WL constraint has been marginalised over a broad range on $h$, whereas the standard sirens one over a narrow range [compare with panels (a) and (b)].
This allows breaking the degeneracy and therefore significantly improving the constraint on $\sigma_8$.
For instance, the error on $\sigma_8$ is $30\%$ from WL alone;
this is reduced to $3\%$ for a joint analysis.
Correlated with $\sigma_8$ and $\Omega_m$ are the $\ln(10^{10}A_s)$ and $n_s$ parameters,
which are respectively constrained to $2\%$ and $7\%$.

\section{Summary}

This paper has set out the general framework with which a joint
analysis of WL and standard sirens could be done in the foreseeable future,
with realistic assumptions on the expected measurement errors.
The benefits of this approach have been illustrated in the figures above.
The improvement of the constraints on dark matter and dark energy, over a WL-only analysis, is evident and always at least an order of magnitude.
The constraints from WL alone would already be competitive with ongoing surveys to date, however with a far smaller number of sources.
The joint analysis with standard sirens would only bring further improvement owing to it adding more information and therefore lifting key degeneracies.
For instance, one of the key findings of this paper is that,
although standard sirens is significantly better than WL in constraining geometry parameters,
the combination of the two observables breaks clustering parameter degeneracies -- in this case, $\Omega_m$ and $\sigma_8$.

Our forecast is based on nominal realistic assumptions about measurement errors, namely luminosity distance, position error, optical redshift, and number density.
We found that shot noise is the limiting factor in our WL analysis, thus not allowing easy access to non-linear scales.
This could in principle be problematic because of the limited available information.
On the other hand, the modelling is generally easier on larger scales, and therefore less prone to potential systematics.

A standard sirens analysis does not require distance ladder calibration -- this dramatically reduces the need of external calibration.
However, given the uncertainties related to the redshift determination,
our joint analysis would probably suffer from systematics that are different from other probes.
Although an accurate assessment of all the systematics and their impact to cosmology is not currently available to date,
the benefit of our approach is obviously clear:
the joint analysis of standard sirens and GW-WL might help solve the tensions in the $H_0$ and $\Omega_m$-$\sigma_8$ spaces between the various cosmological probes,
such as CMB, galaxy WL, and Type Ia SNs, and distinguish between residual systematics and new physics.
We conclude that it is not unrealistic to expect that our joint analysis will probably compete with (if not outperform with increasing number densities) cosmology experiments of the future.

\begin{acknowledgements}
GC acknowledges the following people for fruitful discussion,
interaction, and feedback at different stages of this research:
S.\ Alam, M.\ Cataneo, J.\ Gair, A.\ Hall, C.-A.\ Lin,
L.\ Lombriser, J.\ Peacock, J.~H.\ Deraps,
and all the WL lunch attendees (Edinburgh);
E.\ N.\ Chisari (Oxford);
A.\ Sesana (Birmingham).
AT thanks the Royal Society for the support of a Wolfson Research Merit Award.
\end{acknowledgements}

%%%%%%%%%%%%%%%%%%%%%%%%%%%%%%%%

%\appendix

\bibliography{references}

\end{document}